\documentstyle[multicol,aps,preprint,epsfig,psfig,epsf,pre]{revtex}

\begin{document}

\draft

\title{Numerical calculations of the phase diagram of cubic blue phases in
cholesteric liquid crystals}

\author{A. Dupuis, D. Marenduzzo and J.M. Yeomans}

\address{The Rudolf Peierls Centre for Theoretical Physics, 1 Keble Road,
  Oxford OX1 3NP, England}

\maketitle

\begin{abstract}
We study the static properties of cubic blue phases by numerically minimising
the three-dimensional, Landau-de Gennes free energy for a cholesteric liquid 
crystal close to the isotropic-cholesteric phase transition.
Thus we are able to refine the powerful but approximate, semi-analytic
frameworks that have been used previously.
We obtain the equilibrium phase diagram and discuss it in relation to
previous results. We find that the value of the
chirality above which blue phases appear 
is shifted by 20\% (towards experimentally more accessible regions)
with respect to previous estimates. We also find
that the region of stability of the O$_5$ structure -- which has not been
observed experimentally -- shrinks, while that of BP I (O$_8^-$) increases
thus giving the correct order of appearance of blue phases at small
chirality. We also study the approach to equilibrium starting from the
infinite chirality solutions and we find that in some cases the 
disclination network has to assemble during the equilibration. In these
situations disclinations are formed via the merging of isolated aligned 
defects.
\end{abstract}

\section{Introduction}

Liquid crystals are typically composed of highly anisotropic molecules.
They are viscoelastic materials; some of their properties
are typical of a liquid while others are usually associated with solids. 
As liquids, they can flow and they exhibit a viscous response to
an applied stress. However, liquid crystals also
possess long-range, orientational 
order \cite{degennes,chandrasekar} which results from the entropic
advantage of aligning the constituent molecules.
The long-range, orientational order is usefully described by
the director field $\vec{n}$, the coarse-grained, average, molecular
orientation.  In a cholesteric or chiral nematic liquid crystal
$\vec{n}$ has a natural twist deformation along an axis perpendicular to
the molecules \cite{degennes,chandrasekar}.  
Examples of cholesteric liquid crystals are  
DNA molecules in solution, colloidal suspensions of 
bacteriophages\cite{bacterio}, and solutions of nematic mixtures such as E7 
with chiral dopants which are widely used in display devices.

A particularly intriguing phase of liquid crystals is obtained by
slowly cooling down a liquid crystal  
from the isotropic into the cholesteric phase. 
Instead of a direct transition into a helical configuration, 
it was found experimentally that the system passes through a series of 
first order phase transitions, all of which occur in a temperature range of
roughly $1 K$ \cite{mermin,bpe}. These phases are known as blue phases. 
Typically \cite{mermin}, experiments report a series of at least three
phases intervening in this small temperature range. The
series of transitions is often as follows:
isotropic (I) $\to$ blue phase III (BP III) $\to$ blue phase II
(BP II) $\to$ blue phase I (BP I) $\to$ cholesteric (C). BP I and BP II
display a cubic symmetry, while BP III has an amorphous nature.
Blue phases are now beginning to find applications in lasers 
\cite{lasers} and in electric field driven devices \cite{polymer}.

Blue phases provide a particularly fascinating example of liquid crystal 
ordering as they correspond to complicated director fields which, even
in equilibrium, are threaded by a regular network of disclinations
\cite{hornreich,bphc0,bphc1,bplc1,bplc2,sethna}. 
Identifying their structure presented a considerable theoretical
challenge the resolution of which is clearly summarised in the
review by Wright and Mermin \cite{mermin}.

In the literature attention has mainly focussed on four different 
candidates for the cubic blue 
phases. The nomenclature used here refers to the symmetry group which 
characterises the lattice of disclinations formed in the blue phases
following \cite{mermin,bplc1,bplc2}.
O$_2$ has the symmetry of a simple cubic lattice, 
$O_8^{+,-}$ and O$_5$ that of a body-centered-cubic lattice.
The two O$_8$ phases are candidates for BP I as they have the same
octahedral symmetry, while O$_2$ has been proposed as a model for BP II. 
The defect structure in each phase in shown in Figure 1. In O$_2$ and in 
O$_5$ the defect lines merge in the center of the unit cell, whereas in
the octahedral phases the defects avoid each other. 

The ${\bf Q}$ tensor theory, in which the local conformation of the
liquid crystal is specified by a tensor, and not only by the
director field ${\vec n}$, is the natural language to understand the
equilibrium properties of blue phases because it is able to capture
their inherent biaxiality \cite{mermin}. With the advent of more powerful
computers it is now possible to numerically minimise the free energy
for any value of the chirality $\kappa$ without further approximation
than that inherent in the free energy expression itself. This is the
programme we follow in this paper.

We determine the equilibrium phase diagram which specifies which
of the cubic blue phases has the lowest free energy as 
a function of the system parameters. 
We find a phase diagram qualitatively similar to that found in earlier work 
and compatible with the assignments BP I = O$_8^-$ and BP II = O$_2$.
However there are quantitative differences. In particular the
results show that the triple point 
between the cholesteric and blue phases is lowered in chirality 
by around 20 \%.  
We also show that the region of stability of O$_5$ shrinks and
that of O$_8^-$ increases with respect to the previous 
estimates. The revised phase diagram is more
compatible with observations suggesting that the Landau-de Gennes theory
is sufficient to explain the static observations on blue phases. 
Finally, we characterise the approach to equilibrium and
in particular we describe the dynamic pathway by which the disclination
structure of O$_8^-$ assembles in the simulations.

\section{Landau-de Gennes theory for cholesteric blue phases}

The equilibrium properties of the liquid crystal are described 
by a Landau-de Gennes free energy density expanded in
terms of a tensor order parameter ${\bf Q}$. This is related to the 
direction of individual molecules, ${\hat{n}}$, by $Q_{\alpha\beta}= 
\langle \hat{n}_\alpha \hat{n}_\beta - {1\over 3}\delta_{\alpha\beta}\rangle$
where the angular brackets denote a coarse-grained average and the Greek 
indices label the Cartesian components of ${\bf Q}$.
The tensor ${\bf Q}$ is traceless and symmetric. Its largest 
eigenvalue, $\frac {2} {3} q$, $0<q<1$, describes the 
magnitude of the order. 

The free energy comprises
a bulk term ${\cal F}_b$ (summation over repeated indices is 
implied hereafter in our notation)
\begin{eqnarray}
{\cal F}_{b}=\frac{A_0}{2}(1 - \frac {\gamma} {3}) Q_{\alpha \beta}^2 - 
          \frac {A_0 \gamma}{3} Q_{\alpha \beta}Q_{\beta
          \gamma}Q_{\gamma \alpha} 
+ \frac {A_0 \gamma}{4} (Q_{\alpha \beta}^2)^2
\label{eqBulkFree}
\end{eqnarray}
and a distortion term, ${\cal F}_d$ which for cholesterics is
\cite{degennes,mermin}
\begin{equation}\label{eqDistFree}
{\cal F}_{d} = \frac{K}{2} \left[\left(\partial_\beta 
Q_{\alpha \beta}\right)^2 + \left(\epsilon_{\alpha \zeta\delta }
\partial_{\zeta}Q_{\delta\beta} + {2q_0}
Q_{\alpha \beta}\right)^2 \right]
\end{equation}
where $K$ is an elastic constant and $q_0=2\pi/p$, with  $p$ 
the pitch of the cholesteric
liquid crystal. The tensor $\epsilon_{\alpha \zeta\delta}$ 
is  the Levi-Civita antisymmetric third-rank tensor, $A_0$ is a constant and 
$\gamma$ controls the magnitude of the order (physically
it corresponds to an effective temperature or concentration 
for thermotropic and lyotropic liquid crystal respectively). 
This free energy 
has the same functional form of the one employed in Refs. 
\cite{bphc0,bphc1}, but the latter is more general as it contains one more
parameter in the bulk free energy density term. 

Of particular interest for the present discussion of blue phase 
equilibrium \cite{mermin} are two quantities,
the chirality $\kappa$, and the reduced temperature
$\tau$. These are defined in Ref. \cite{bphc1} and 
in our formulation they can be found via:
\begin{eqnarray}
\label{transformation}
\kappa & = & \sqrt{\frac{108 Kq_0^2}{A_0\gamma}},\\ \nonumber
\tau & = &\frac{27 A_0(1-\gamma/3)+108 Kq_0^2}{A_0\gamma}=
\frac{27(1-\gamma/3)}{\gamma}+\kappa^2.
\end{eqnarray}

Approaches to study the equilibrium properties of blue phases 
which have given a great deal of insight into the phase behaviour
have so far been based on an expansion in the parameter $\kappa$. 
{\it High chirality} theories are strictly
valid for infinite $\kappa$. In that case an infinite number of exact
minimizers of the free energy density can be found as an arbitrary
sum of biaxial helices \cite{mermin}. 
{\it Low chirality} theories have also been formulated.
These rely on the Frank free energy defined in terms of the
coarse grained director field $\vec{n}$ instead of 
${\bf Q}$: this is equivalent to assuming a uniaxial order parameter.

In this work we do not make assumptions about the value of $\kappa$ but 
rather minimise the Landau-de Gennes free energy by
numerically solving an equation of motion for {\bf Q} \cite{beris}
\begin{equation}
\frac{\partial {Q_{\alpha\beta}}}{\partial t}=\Gamma {H_{\alpha\beta}}
\label{Qevolution}
\end{equation}
where $\Gamma$ is a collective rotational diffusion constant.

The term on the right-hand side of Eq. (\ref{Qevolution})
describes the relaxation of the order parameter towards the minimum of
the free energy. The molecular field ${\bf H}$ which provides the driving
force is related to the derivative of the free energy ${\cal F}$ by
\begin{equation}
{\bf H}= -{\delta {\cal F} \over \delta {\bf Q}}+({\bf
    I}/3) Tr{\delta {\cal F} \over \delta {\bf Q}}
\label{molecularfield}
\end{equation}   
where Tr denotes the tensorial trace. From Eqs. (\ref{eqBulkFree})
and (\ref{eqDistFree}), the molecular field is explicitly 
\begin{eqnarray}
H_{\alpha\beta} & = & \left(A_0(1-\gamma/3)+4Kq_0^2\right)Q_{\alpha\beta}
-A_0\gamma\left(Q_{\alpha\gamma}Q_{\gamma\beta}-
\frac{\delta_{\alpha\beta}}{3}Q_{\gamma\delta}^2\right)\\ \nonumber
& + & A_0\gamma Q_{\gamma\delta}^2 Q_{\alpha\beta}
+K\partial_{\gamma}^2 Q_{\alpha\beta}
-4Kq_0\epsilon_{\alpha\gamma\delta}\partial_{\gamma}Q_{\delta\beta}.
\end{eqnarray}

\section{Numerical algorithm}

Our aim in this work is to numerically minimise -- by solving Eq. 
(\ref{Qevolution}) using a lattice Boltzmann algorithm -- the 
free energies of the cubic blue phases $O_2$, $O_5$, $O_8^{+}$ and $O_8^{-}$.
In this Section we report the procedure used. 

The equilibrium state for chosen values of chirality $\kappa$ and reduced 
temperature $\tau$ was obtained as the steady state solution of the equation 
of motion (\ref{Qevolution}). The initial condition for each phase was taken 
as its configuration for infinite chirality ($\kappa=\infty$ or $A_0=0$) for 
which exact analytical expressions are available \cite{mermin}. In the
$\kappa=\infty$ limit the ${\bf Q}$ tensors characterising the blue phases 
retain the correct topology of the disclination lines. As expected this was 
preserved under the dynamics allowing us to minimise the free energy for a 
structure with the chosen network of disclinations for any value of $\tau$ 
and $\kappa$.

For completeness we report here the initial configurations chosen in this 
work. (Note that in all cases the components $yy$, $zz$, $xz$ and $yz$ 
are obtained by cyclic permutation from those given below.)

\noindent The initial configuration for O$_2$ is
\begin{equation}\label{O2}
Q_{xx} = A \left\{\cos{(2q_0z)}-\cos{(2q_0y)}\right\}, \qquad 
Q_{xy} = - A \sin{(2q_0z)} 
\end{equation}
where $A>0$ is an arbitrary amplitude.

\noindent That for  $O_5$ is
\begin{eqnarray}\label{O5}
Q_{xx} & = & A \left\{2\cos{(\sqrt{2}q_0y)}\cos{(\sqrt{2}q_0z)}
\right . \\ \nonumber
& - & \left. \cos{(\sqrt{2}q_0x)}\cos{(\sqrt{2}q_0z)}
-\cos{(\sqrt{2}q_0x)}\cos{(\sqrt{2}q_0y)}\right\}, \\ \nonumber
Q_{xy} & = & A \left\{\sqrt{2}\cos{(\sqrt{2}q_0y)}\sin{(\sqrt{2}q_0z)}
\right . \\ \nonumber
 & - & \left . \sqrt{2}\cos{(\sqrt{2}q_0x)}\sin{(\sqrt{2}q_0z)}
-\sin{(\sqrt{2}q_0x)}\sin{(\sqrt{2}q_0y)}\right\} \\ \nonumber
\end{eqnarray}
where $A>0$.

\noindent Those for $O_8^{+,-}$ are
\begin{eqnarray}\label{O8}
Q_{xx} & = & A \left\{-2\cos{(\sqrt{2}q_0y)}\sin{(\sqrt{2}q_0z)} \right.\\ 
\nonumber
& + & \left. \sin{(\sqrt{2}q_0x)}\cos{(\sqrt{2}q_0z)}
+\cos{(\sqrt{2}q_0x)}\sin{(\sqrt{2}q_0y)}\right\} \\ \nonumber
Q_{xy} & = & A \left\{\sqrt{2}\cos{(\sqrt{2}q_0y)}\cos{(\sqrt{2}q_0z)}
\right. \\ \nonumber
& + & \left. \sqrt{2}\sin{(\sqrt{2}q_0x)}\sin{(\sqrt{2}q_0z)}
-\sin{(\sqrt{2}q_0x)}\cos{(\sqrt{2}q_0y)}\right\} \\ \nonumber
\end{eqnarray}
where $A$ is positive for $O_8^{+}$ and negative for $O_8^{-}$.

In general, the minimum free energy for blue phases is not
attained when the periodicity of the disclination lattice 
matches the pitch of the cholesteric helix $p$
but when the unit cell is larger.
This is accounted for by considering alternative initial conditions with 
$q_0$ substituted by $rq_0$ \cite{hornreich,bphc0,bphc1}, with $r<1$.
Since the lattice periodicity is bigger than that for the infinite
chirality solutions, $r$ is referred to as a `redshift'. 
We checked the validity of the values of $r$ suggested in 
Ref. \cite{bphc1} and then used them to build the 
initial conditions. If redshift is not 
accounted for, the region of stability of blue phases does not change 
significantly, but the O$_8^-$ phase is not found.

Details of the lattice Boltzmann algorithm used to solve the  
equation of motion are given in Refs. \cite{lblc1,lblc2,chol1}. 
Calculations were performed on a parallel machine, for a 32x32x32
cubic lattice. This means that, for example, in the case of $O_2$ a 
cholesteric pitch was discretized into 64 lattice points. 
Typically, the simulation of one point in the phase space $(\kappa,\tau)$
ran on eight processors and required $8$ hours of computational time.
Periodic boundary conditions were used throughout.
To ensure that equilibrium was reached, we required that the 
variation in the free energy after 200 iterations 
of Eq. (\ref{Qevolution}) was smaller than $10^{-4}$ before ending the run. 
A convenient value of $\Gamma$, ensuring numerical stability and fast 
computation time, was $0.33775$. 

The parameters defining the free energy
were chosen to match typical values for cholesterics.
In what follows the points on the phase diagram will
be labelled by their $(\kappa,\tau)$ (as in Ref. \cite{bphc1})
or $(A_0,\gamma)$ values. These can be related straightforwardly
via the transformation in Eq. (\ref{transformation}).

\section{Results}

In this Section we report the results obtained 
for the defect structure of the blue phases, and the equilibrium 
phase diagram in the vicinity of the isotropic-cholesteric
transition. We also consider the dynamics of the defect lines as equilibrium
is approached.

\subsection{Defect structure: $O_2$, $O_5$, $O_8^{+}$ and $O_8^{-}$}

\begin{figure}
\begin{center}
\begin{tabular}{cc}
(a) O$_2$ & (b) O$_5$ \\
\psfig{figure=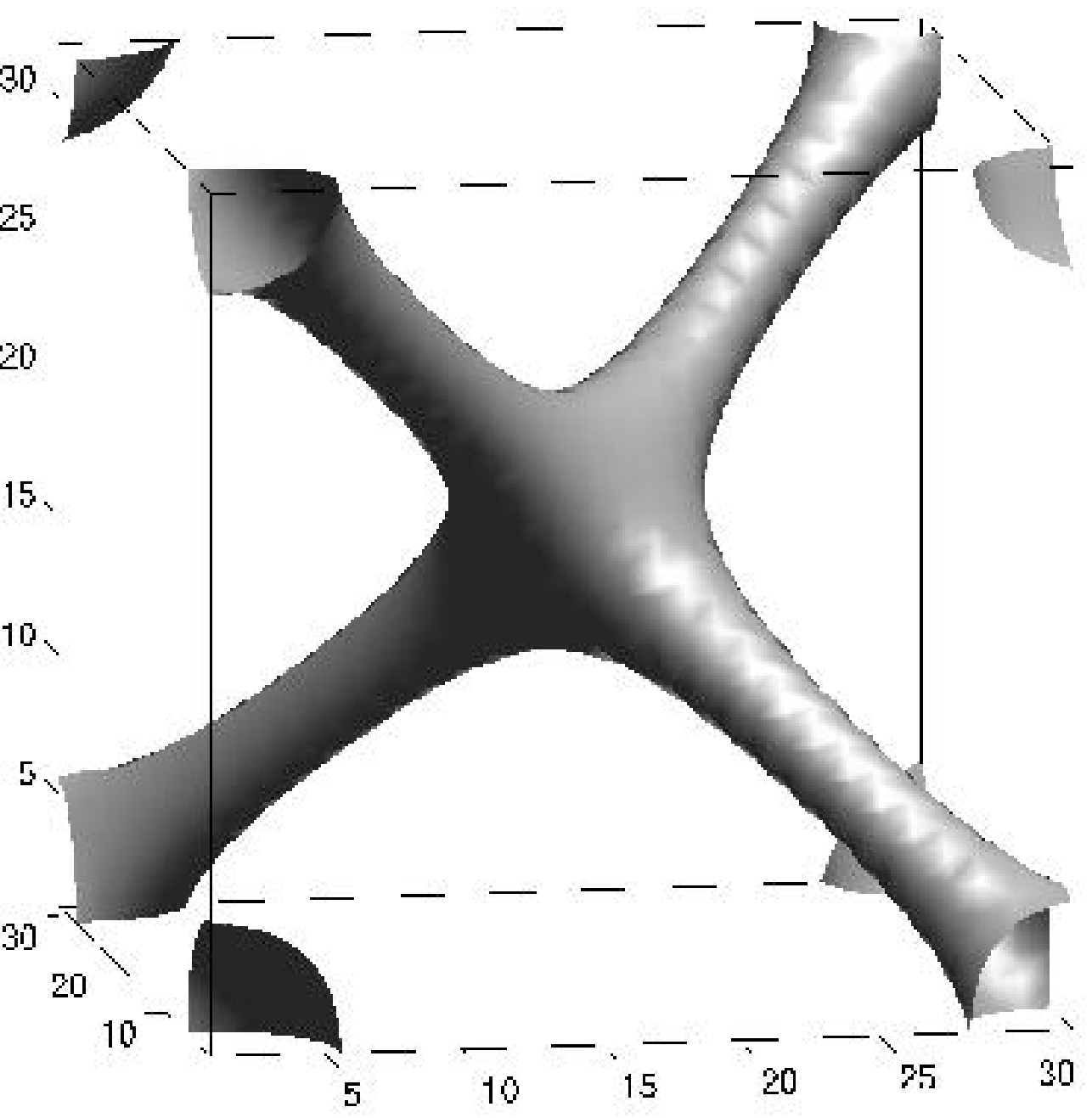,width=5.5cm} &
\psfig{figure=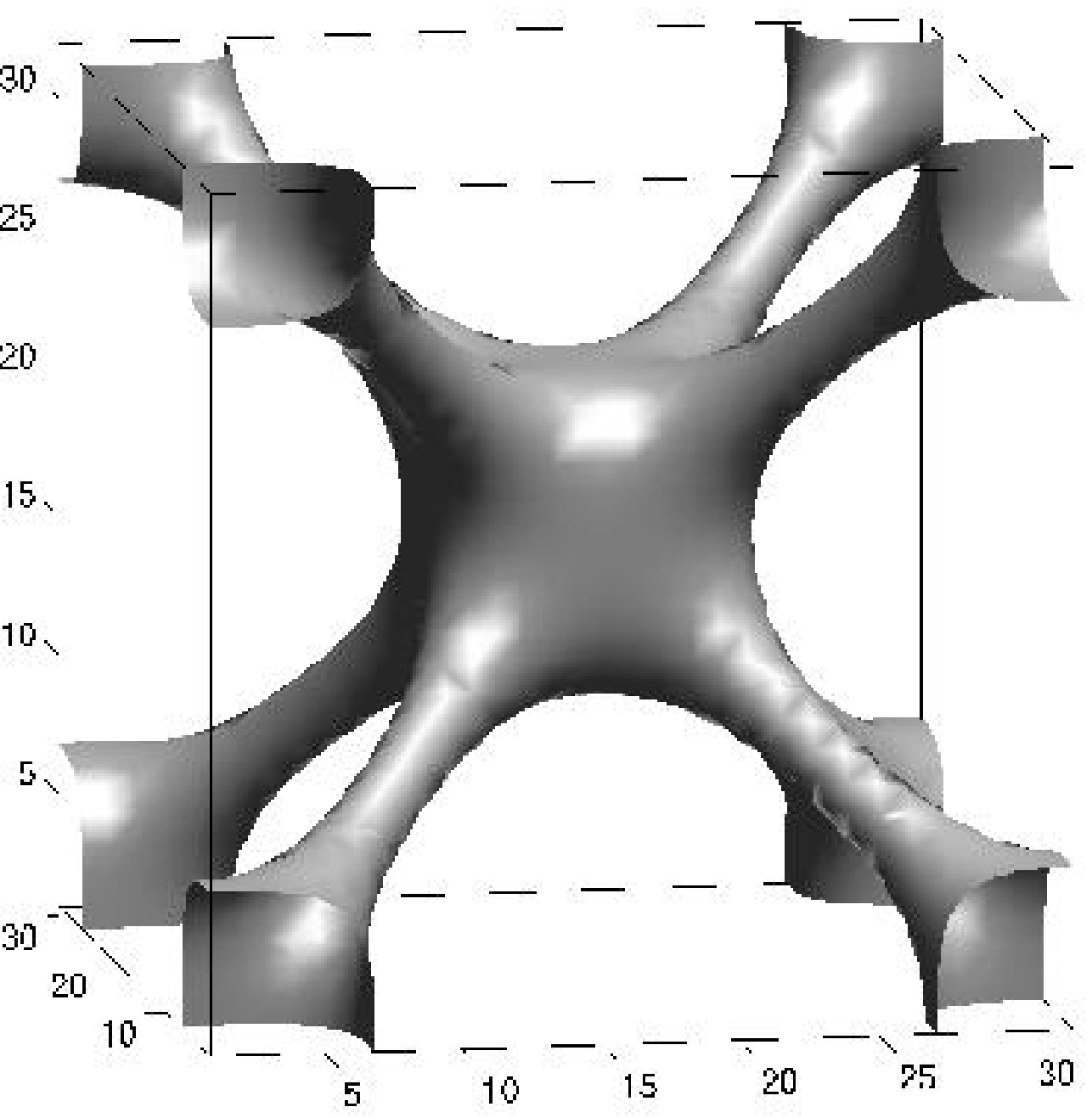,width=5.5cm} \\
(c) O$_8^+$ & (d) O$_8^-$ \\
\psfig{figure=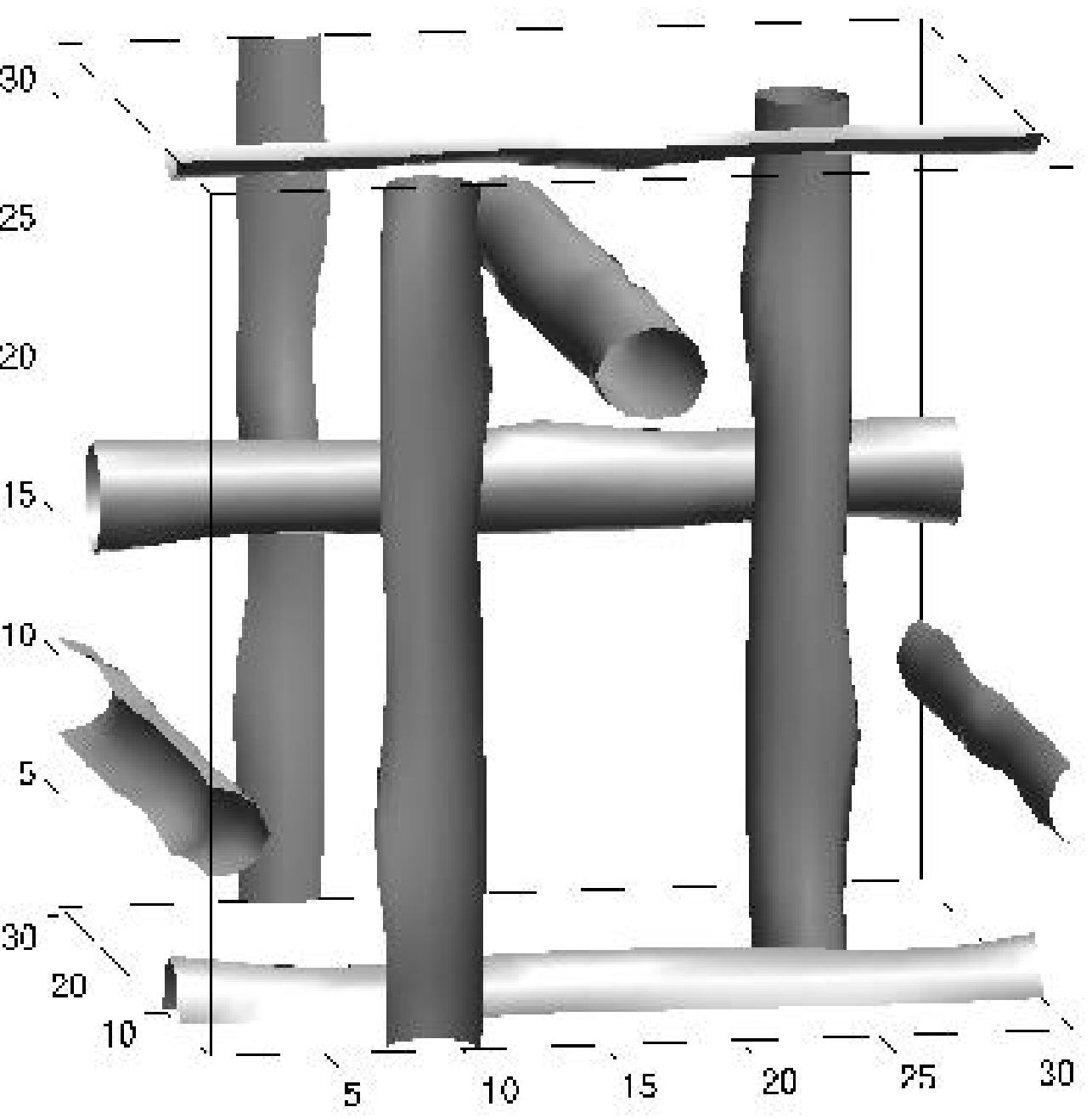,width=5.5cm} &
\psfig{figure=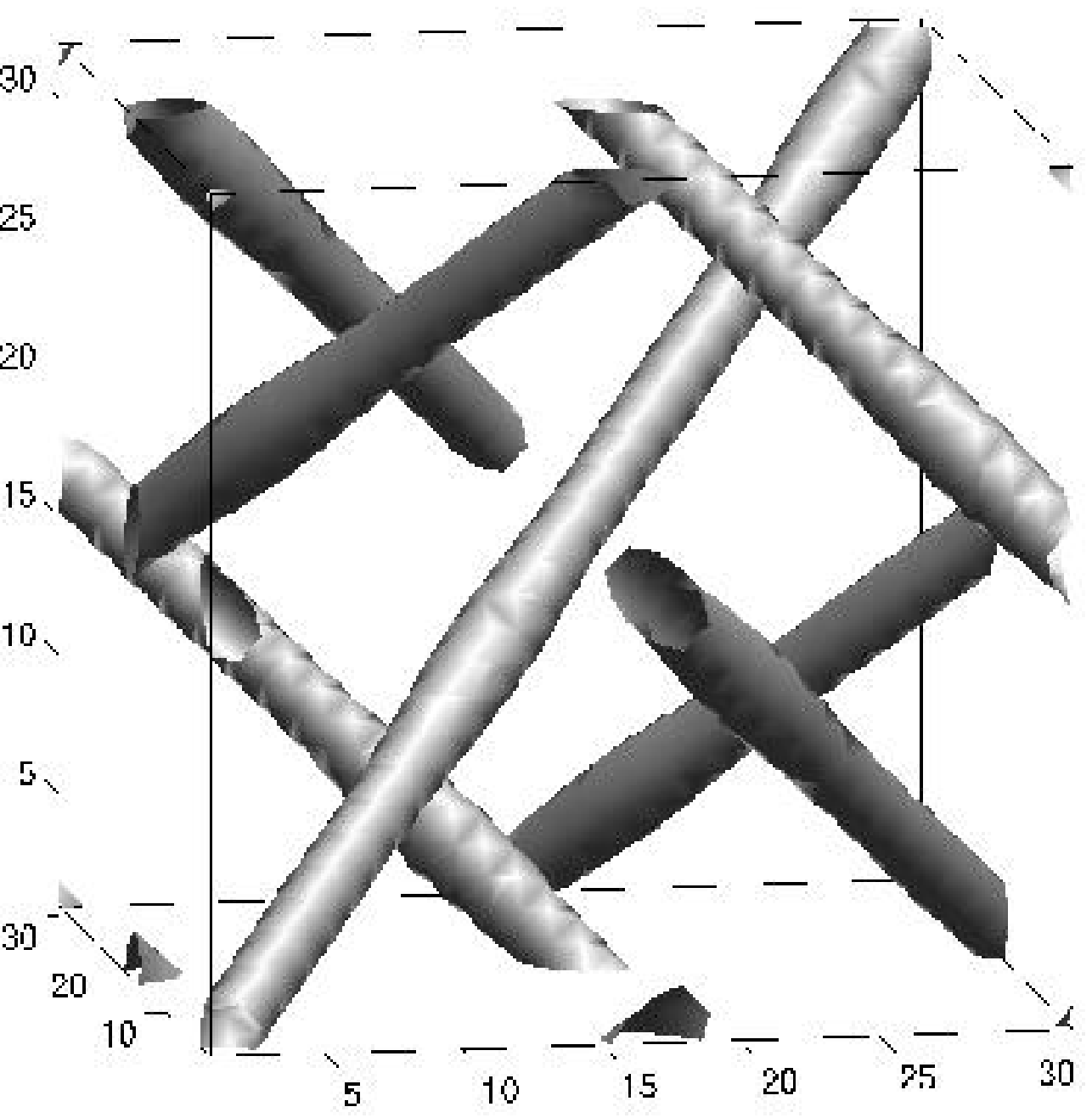,width=5.5cm} \\
\end{tabular}
\end{center}
\caption{Defect structure where $\gamma=2.80$ and $A_0=0.001$
for (a) O$_2$, (b) O$_5$, (c) O$_8^+$ and (d) O$_8^-$. }
\label{defect_structure}
\end{figure}

As a first check, we compute the defect structure associated with the
cubic blue phases after equilibration. We choose $A_0=0.001$ and
$\gamma=2.8$, while the elastic constant $K=0.01$ for $O_2$, and
$K=0.005$ for $O_5$ and $O_8^{+,-}$. These correspond to a chirality
$\kappa=1.93$ and a reduced temperature $\tau=4.37$.
The resulting disclination line networks are shown in Fig.
\ref{defect_structure}. The tubes in the figure represent regions of
the blue phase unit cell in which the order parameter
drops below some specified threshold, i.e. they represent disclination lines.
The thickness of the tubes is related to the width of 
the defect cores.

These structures are in good agreement with those obtained in Refs. 
\cite{bphc1,bplc1,bplc2}. We note that the structure of the defects 
depends on the
parameters. If $A_0$ or $\gamma$ are increased, i.e. if $\kappa$ decreases,
then the width of the disclinations decreases and the drop in order parameter
at the defect cores becomes shallower. 

We also checked the three-dimensional director
field profile and found the presence of distinct regions of double twist 
separated by the defects shown in Fig. \ref{defect_structure}, in 
qualitative agreement with the usual theoretical picture of blue 
phases~\cite{bphc0,bphc1,sethna}.

\subsection{Phase diagram}

\begin{figure}
\begin{center}
\psfig{figure=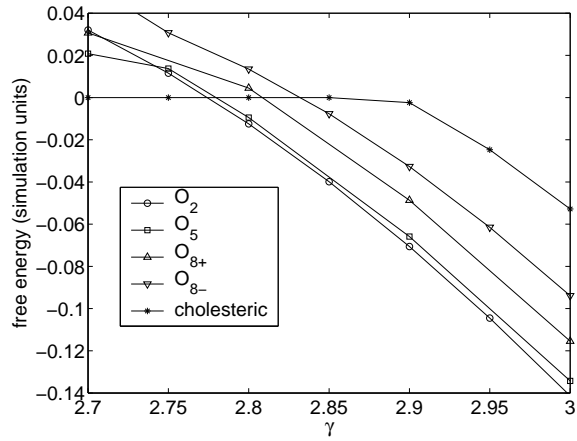,width=3.in}
\end{center}
\caption{Free energy of O$_2$, O$_5$, O$_8^+$, O$_8^-$ and 
the cholesteric helix
as a function of $\gamma$ for $A_0=0.001$. }
\label{free_energy_plots}
\end{figure}

Next we report results for the free energy of the equilibrated blue phases. 
In Fig. \ref{free_energy_plots} we show free energy curves
for O$_2$, O$_5$, O$_8^+$, O$_8^-$ and the isotropic and cholesteric phases
for $A_0=0.001$ and variable $\gamma$. It can be
seen that blue phases appear near the transition to the isotropic phase. 
As $A_0$ increases, the region of stability of blue phases 
shrinks. This is in agreement with previous experimental and analytical 
results in the literature \cite{bpe,bphc1,bplc1,bplc2}.

\begin{figure}
\centerline{
\psfig{figure=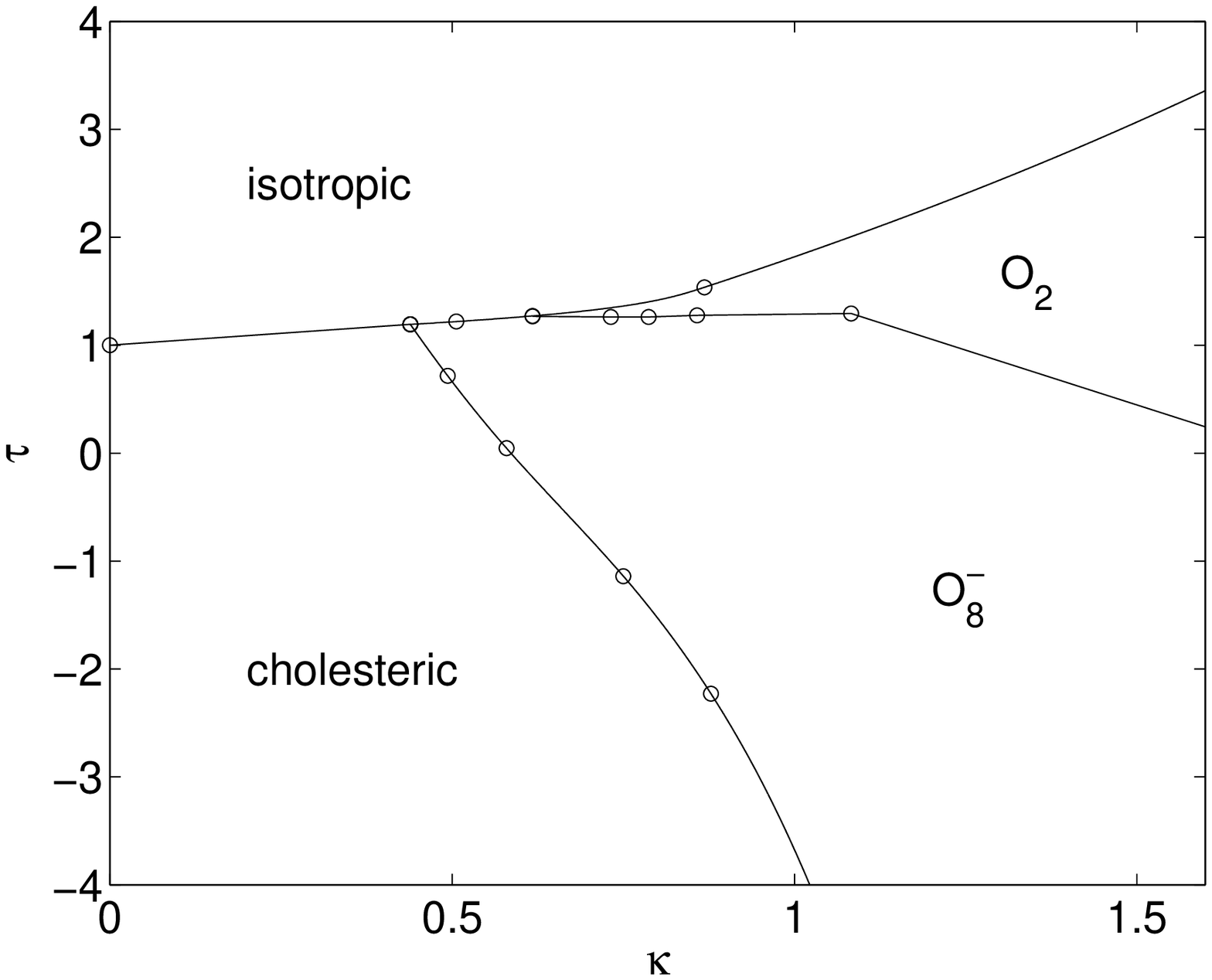,width=8.cm}
\psfig{figure=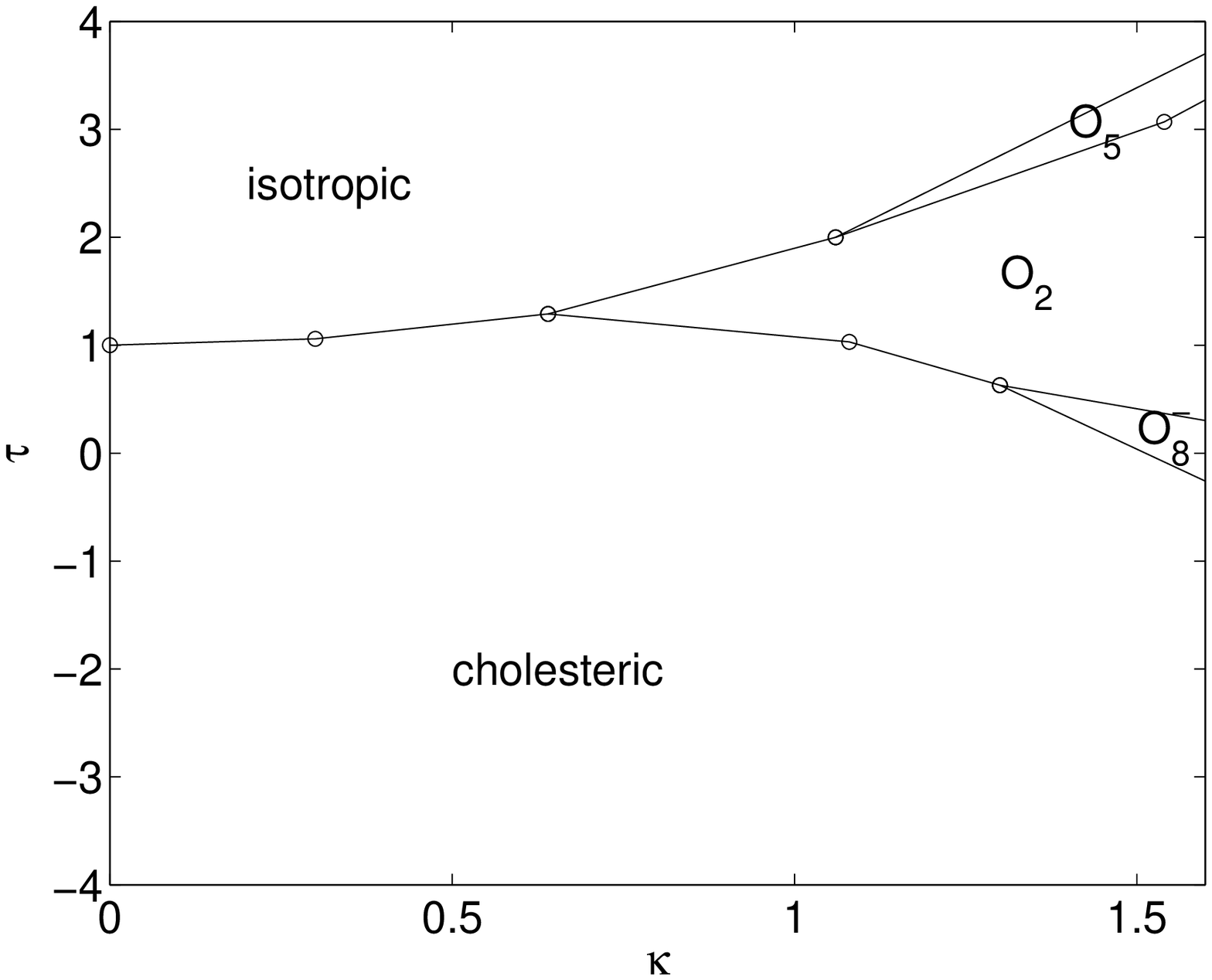,width=8.cm}}
\caption{{Left, phase diagram in the $(\kappa,\tau)$ plane obtained
numerically (this work). Right, phase diagram from Ref. [10].}}
\label{phase_diagram}
\end{figure}

In Fig. \ref{phase_diagram}, we show the phase diagram in the 
$(\kappa,\tau)$ plane which
identifies for every point which of the phases,
$O_2$, $O_5$, $O_8^{+,-}$, cholesteric or isotropic, has the minimum free
energy. For comparison we 
show, in the same plane, the phase diagram obtained in Ref. \cite{bphc1} 
which employed a Fourier expansion of the tensor order parameter ${\bf Q}$
around the $\kappa=\infty$ limit.
We stress that the main difference is that in our case we relax the
structures (with the disclination network topology suggested in Ref. 
\cite{bphc1}) by dynamically solving Eq. (\ref{Qevolution}). So the 
configurations used to construct our phase diagram are actual (in general 
local) minima of the Landau-de Gennes free energy.
In Ref. \cite{bphc1}, the configurations found 
correspond to the free energy minima within a restricted phase space in which 
the ${\bf Q}$ tensor expression is constrained to be a sum of
spherical harmonics of order $m\le 2$ with variable coefficients.

Overall, we find qualitative agreement in that (a) the stable phases found 
(for the parameters investigated here) 
are O$_2$, O$_5$ and O$_8^-$, (b) O$_5$ appears for high chirality only, and
(c) the phase diagram is consistent with the assignment of BP I as O$_8^-$
and of BP II as O$_2$ proposed in \cite{bphc1,bplc2}. However, a few relevant
differences should be stressed. First, the region of stability of O$_5$
is shifted to unphysically high values of the chirality $\stackrel
{>}{\sim}2$. 
(Values of $\kappa\stackrel{<}{\sim}1.2$ are found in
cholesterics.) This result is consistent with
the observations, which ruled out structures of O$_5$ symmetry from the
known blue phases. The region of stability of O$_8^-$
expands and reaches lower chirality values. The critical point below which
blue phases are no longer stable is also shifted from $\kappa\sim 0.6$
to below $\kappa=0.5$ (deeper into the experimentally accessible region).
Pleasingly, this leads to the correct order of appearance of BP I and BP II 
as the chirality is increased (as found in the experiments).
This was not achieved before within the Landau-de Gennes framework.

\begin{figure}
\centerline{\psfig{figure=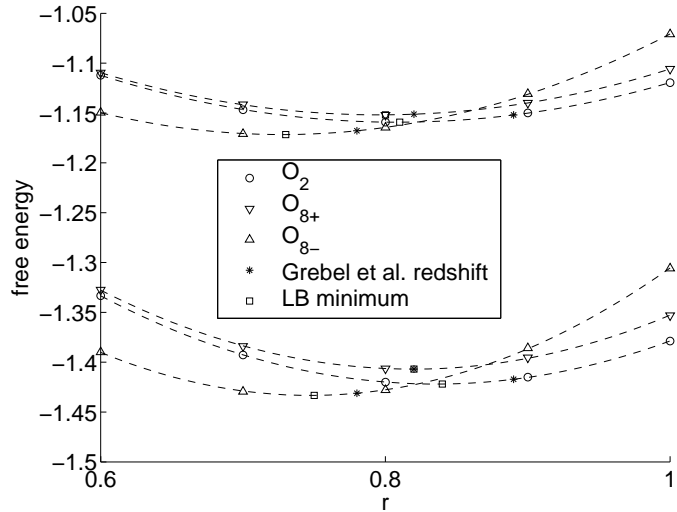,width=9.cm}}
\caption{Plot of the free energies (simulation units) 
of O$_2$, O$_8^+$, O$_8^-$ for two
points in the $(A_0,\gamma)$ plane ((0.006,3) for the top three curves 
and (0.002,3.5) for the bottom three). Dashed lines are quadratic fits. 
Squares and stars show the numerical minimum and the one
found in Grebel et al. [10] respectively. }
\label{redshift}
\end{figure}

To construct the phase diagram in Fig. \ref{phase_diagram}, we used
the values of the redshift from Ref. \cite{bphc1}, which were
obtained within the approximate framework detailed above. As a check, for a 
few points in the phase diagram, we equilibrated systems with different
redshifts. Plots of free energy versus redshifts for two points in
the phase diagram are shown in Fig. \ref{redshift}.
At all points we tested we found that the redshift values in Ref. 
\cite{bphc1} corresponded very closely to the free energy minimum.
In this way we feel confident that an extensive calculation
taking into account all possible redshifts for all data points, 
which is numerically extremely expensive, would give a
quantitatively very similar phase diagram. 

The discrepancy between the phase diagram obtained here and the one
given in Ref.~\cite{bphc1} show that high order 
Fourier components of ${\bf Q}$
are important at least in some regions of the phase diagram. Indeed the 
authors of Ref.~\cite{bphc1} noted that in a large part of the 
region where the phase diagrams differ the energies of the various phases 
are very close. Thus  considering
a larger variational space could well change the phase boundaries.

\subsection{Dynamics of the approach to equilibrium}

It is interesting to consider the dynamics of the approach 
to equilibrium of the blue phases from the infinite chirality
initial states. The O$_2$ and O$_8^-$ structures are initially 
quite far from their
equilibrium configurations, while the 
O$_5$ and O$_8^+$ configurations are closer. This means that  
the free energy for O$_5$ is lower than that of O$_2$ or O$_8^-$ at
the beginning of the simulations. During equilibration, the finite chirality
deforms the tensor configurations of  O$_2$ 
and O$_8^-$ via a process which can be suggestively compared to the
acquisition of an infinite number of harmonics in the language of  
\cite{bphc1} and the curves eventually cross. The approach to
equilibrium of the free energy was in all cases found to be well 
represented by an exponential decay to the equilibrated value,
with the decay rate playing the role of a relaxation time. Since the
transitions are all (weakly) first order this is as expected on
general grounds.

\begin{figure}
\begin{center}
\centerline{
\psfig{figure=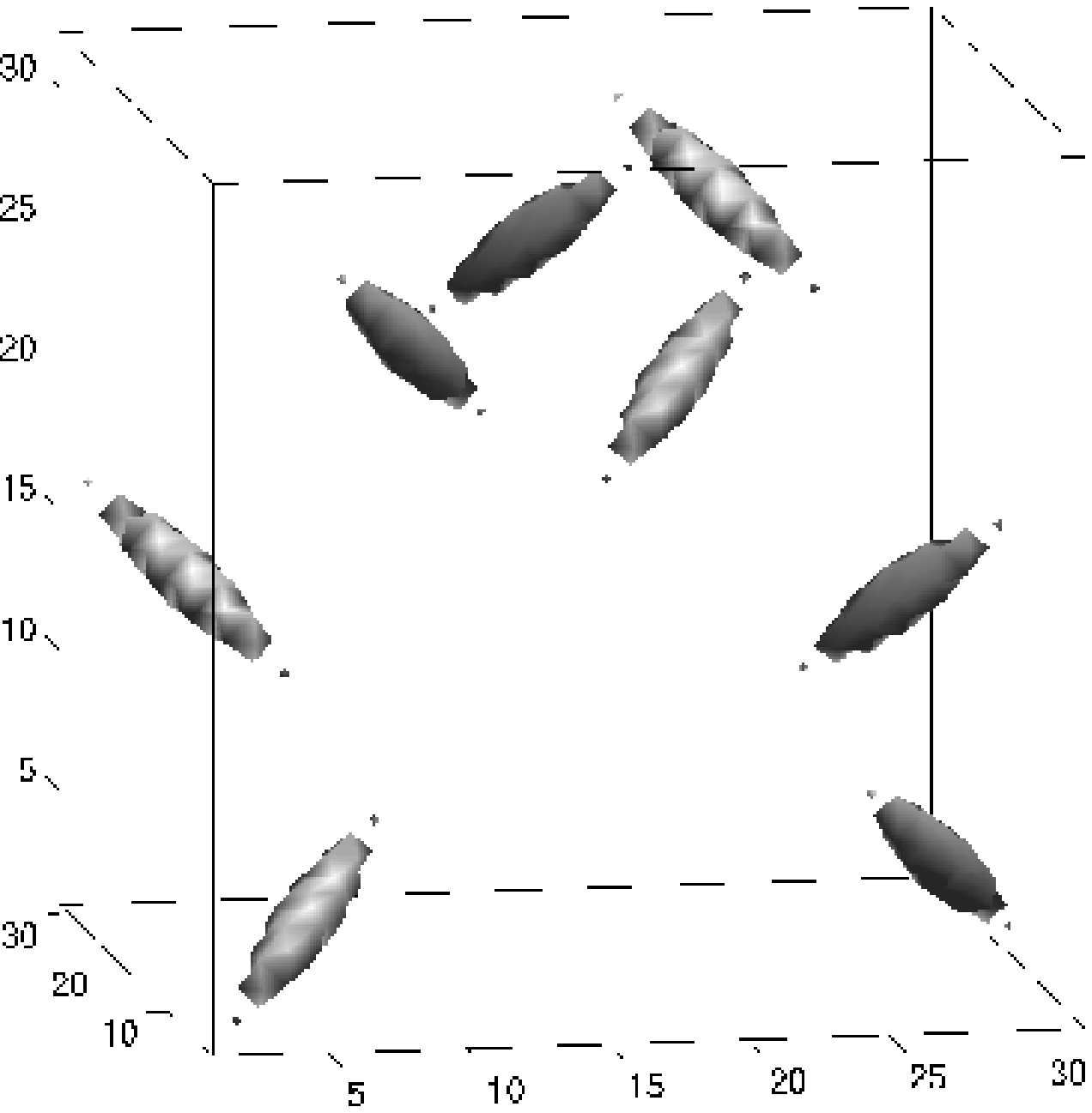,width=3.1cm}
\psfig{figure=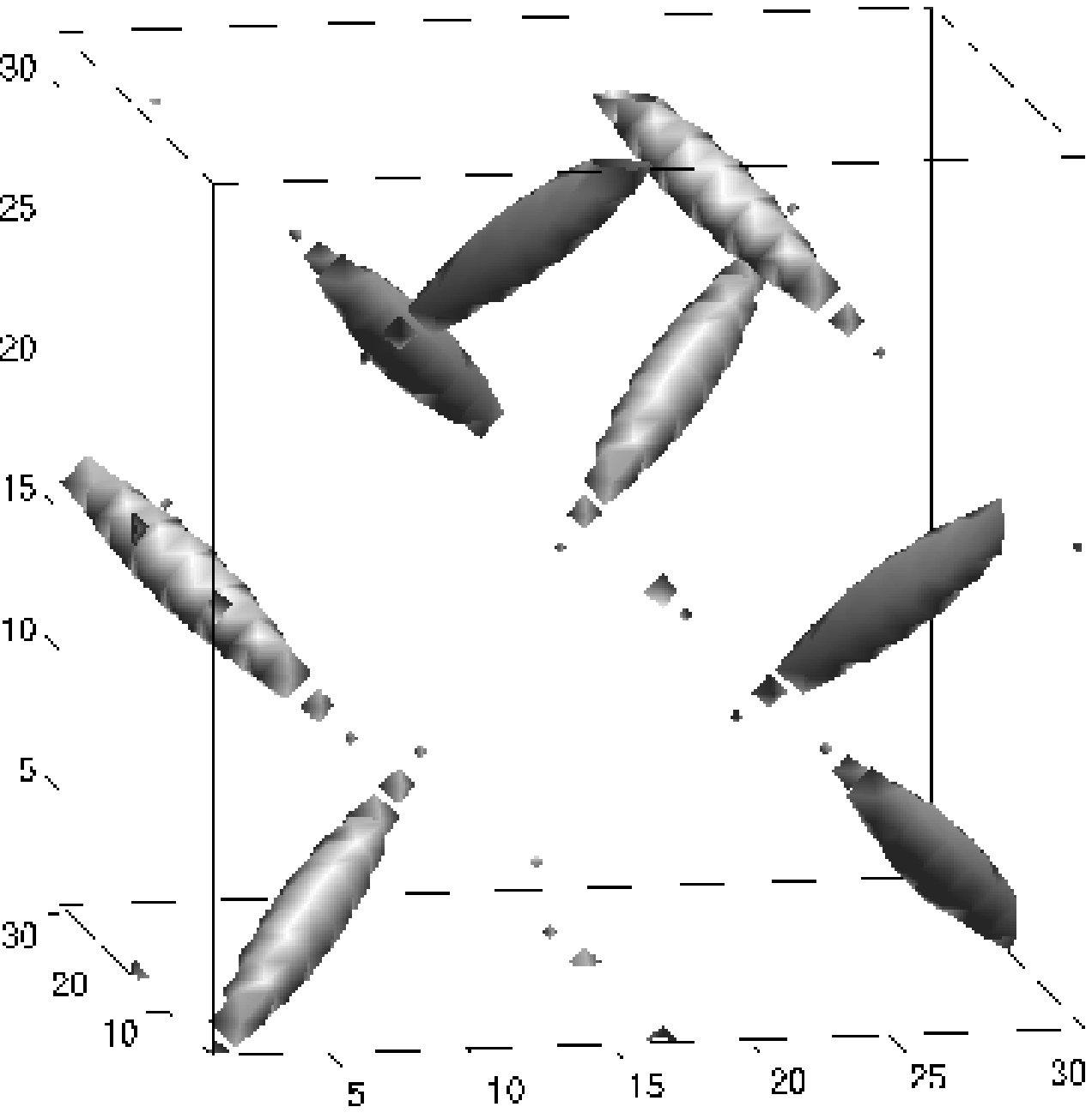,width=3.1cm}
\psfig{figure=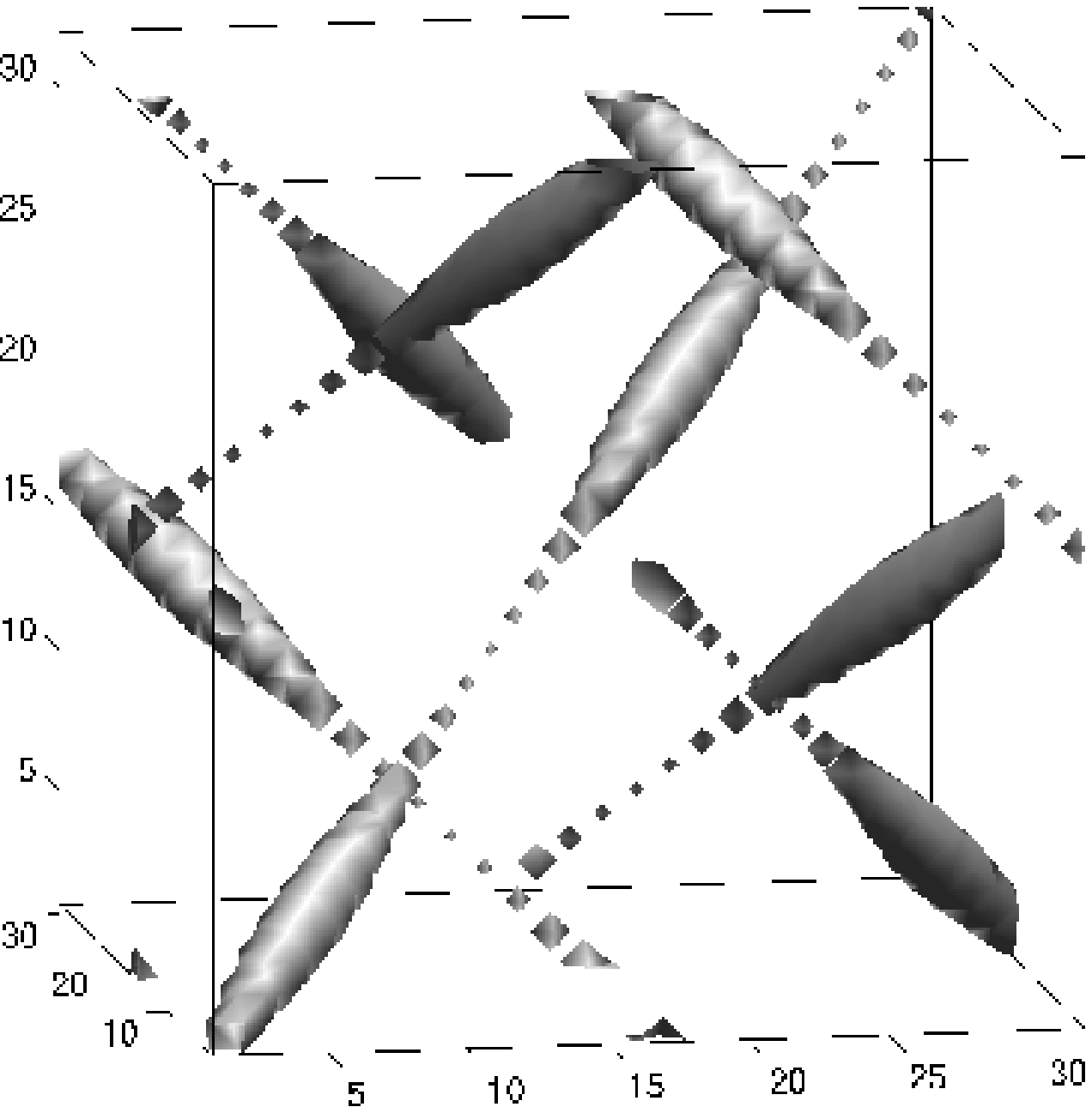,width=3.1cm}
\psfig{figure=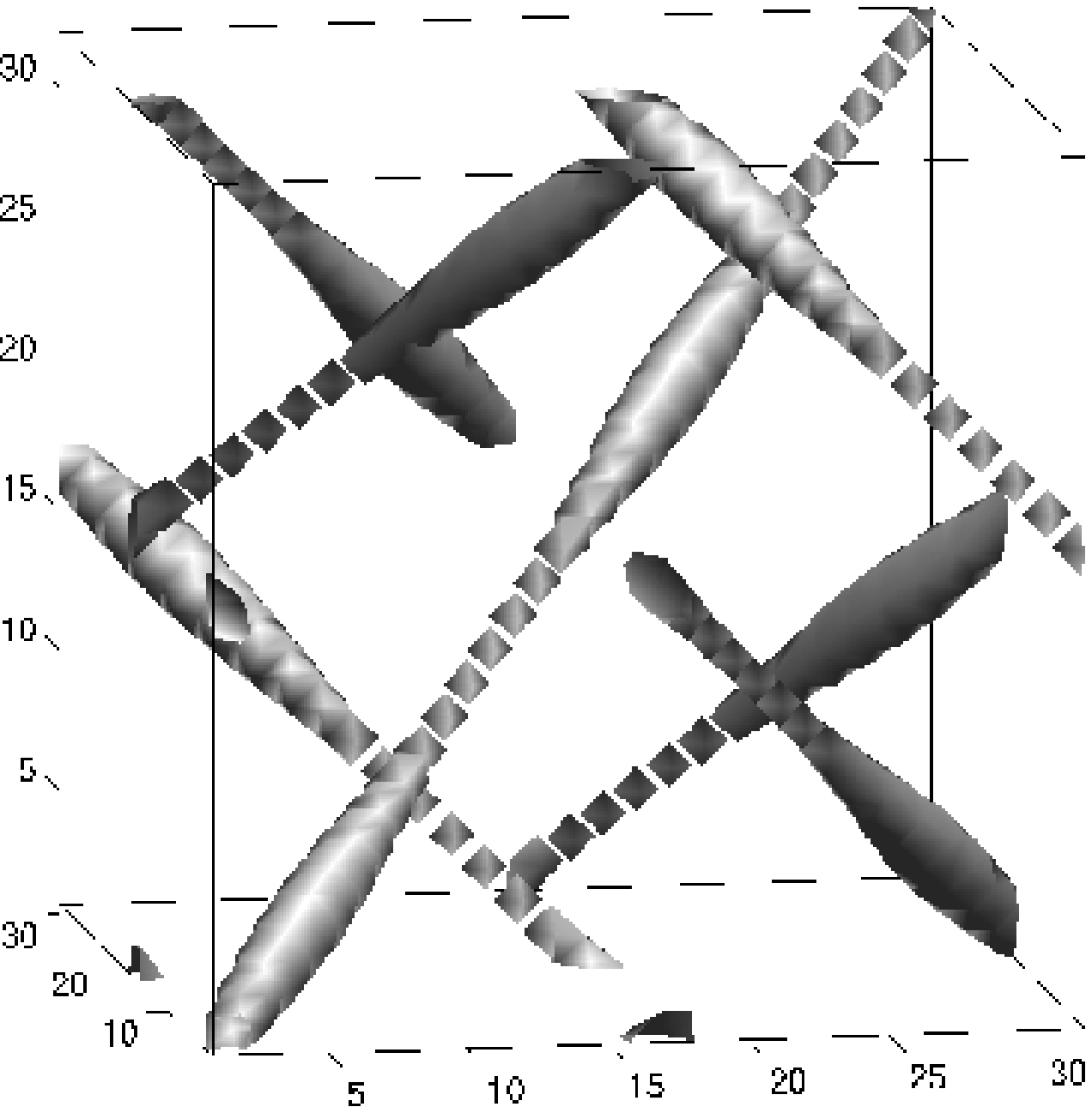,width=3.1cm}
\psfig{figure=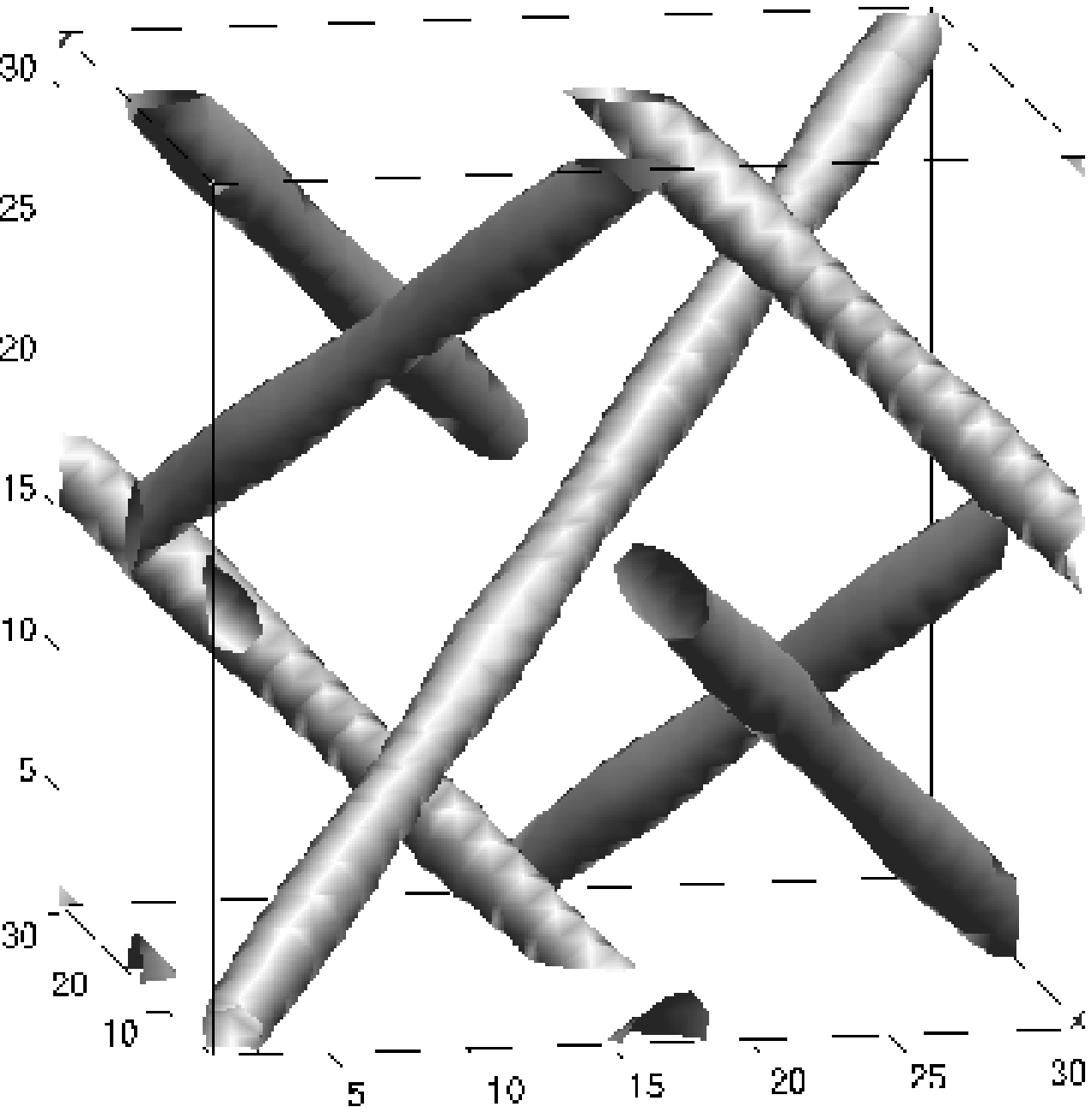,width=3.1cm}}
\end{center}
\caption{Evolution of the defect structure (from left to right)
for $\gamma=2.80$ and $A_0=0.001$,
for the $O_8^-$ configuration. The initial configuration is stable for
$A_0=0$.}
\label{dynamicsO8-}
\end{figure}

The most interesting dynamics of approach to equilibrium is that of
$O_8^-$. In this case the initial structure does not closely resemble the
equilibrium structure (see Fig. \ref{dynamicsO8-}). 
In particular the disclination line along the axis $\left [ 111 \right ]$ is 
only partially present. During equilibration, it assembles by the
formation of isolated local defects aligned along $\left [ 111 \right ]$
which then merge into a disclination line. 
The dynamical evolution was found to be almost independent of
the values of $A_0\ne0$ and $\gamma$ studied. It would be interesting to 
follow the evolution of the disclination line experimentally to
see whether this effect can be realistic. Calculations are currently
underway to check if the same pathway is followed when the full
hydrodynamic equations of motion of the liquid crystal are solved.

\section{Conclusions}

In conclusion, we have numerically minimised the Landau-de Gennes free energy
of a cholesteric liquid crystal and identified the region of stability of the
blue phases. It was possible to obtain results for any values of the chirality
$\kappa$, that is, for any degree of biaxiality of the tensor order
parameter. Our results qualitatively confirm those obtained previously
using expansions for large or small $\kappa$. However there are
quantitative differences; in particular the blue phases first appear at
a value of $\kappa$ $\sim$ 20\% smaller than in the approximate
minimisations. Also, it is 
interesting to note that the full numerical solution presented here 
produces for the first time the correct order of appearance of BP I and BP II
at low chiralities, while the O$_5$ phase, which was not observed
in experiments, is relegated to unphysical values of the chirality. We 
finally followed the approach to equilibrium. In the case of O$_8^-$
the disclination network assembles during the equilibration via an
intermediate state in which aligned isolated defects appear and
then join to form the disclination line along the cube lattice diagonal.

Note that in two rather recent papers \cite{fluctuations1,fluctuations2},
it was found that adding fluctuations to the Landau-de Gennes free
energy could render the phase diagram more realistic in that e.g. the
region of stability of O$_5$ was restricted. Here we find a similar
result by an exact minimisation of the mean field free energy
without including fluctuations.

Finally, our approach can be generalised to study the effect of
an electric field on the phase diagram as well as the 
dynamics of the switching of a blue phase device, such as the one
proposed in Ref.~\cite{polymer}. 
Further work is underway along these lines.\\

This work was supported by EPSRC grant no. GR/R83712/01 and by EC IMAGE-IN 
project GRD1-CT-2002-00663. We thank the Oxford Supercomputing Center and 
the Scientific and Parallel Computing group from the University
of Geneva for providing supercomputing resources. We are grateful to 
E. Orlandini for many important discussions.

\end{document}